\documentclass[aps,prl,reprint,twocolumn,superscriptaddress,showpacs]{revtex4-2}

\usepackage{graphicx}
\usepackage{mathrsfs}
\usepackage{bm}
\usepackage{amsmath}
\usepackage{dcolumn}
\usepackage{dsfont}
\usepackage{amssymb}
\usepackage{tabularx}
\usepackage{array}
\usepackage{float}
\usepackage{color}
\usepackage{epstopdf}
\usepackage{mathrsfs}
\usepackage[T1]{fontenc}
\usepackage[colorlinks, linkcolor=blue,anchorcolor=blue,citecolor=blue,urlcolor=blue]{hyperref}

\usepackage[mathlines]{lineno}
\usepackage{bbold}
\usepackage[T1]{fontenc}
\usepackage{multirow}

\usepackage{gensymb}
\graphicspath{{./}{figure/}}

\begin{document}

\preprint{\textit{Preprint: \today, \now.}} %For internal use only, do not distribute.}}

\title{Electronic Correlations and Absence of Superconductivity in the Collapsed Phase of LaFe$_2$As$_2$}

\author{Jianzhou Zhao}\email{jzzhao@swust.edu.cn}
\affiliation{Co-Innovation Center for New Energetic Materials, Southwest University of Science and Technology, Mianyang, 621010, People's Republic of China}
\affiliation{Research Laboratory for Quantum Materials, Singapore University of Technology and Design, Singapore 487372, Singapore}

\author{Yilin Wang}\email{yilinwang@ustc.edu.cn}
\affiliation{Hefei National Laboratory for Physical Sciences at Microscale, University of Science and Technology of China, Hefei, Anhui 230026, China}

\author{Xiaolong Feng}
\affiliation{Research Laboratory for Quantum Materials, Singapore University of Technology and Design, Singapore 487372, Singapore}

\author{Shengyuan A. Yang}
\affiliation{Research Laboratory for Quantum Materials, Singapore University of Technology and Design, Singapore 487372, Singapore}
\affiliation{Center for Quantum Transport and Thermal Energy Science, School of Physics and Technology, Nanjing Normal University,
Nanjing 210023, China}

%
%============== Abstract ==================%
\begin{abstract}
The interplay between structural phase and electronic correlations has been an intriguing topic of research. An prominent example is the pressure-induced uncollapsed to collapsed tetragonal phase transition observed in CaFe$_2$As$_2$, which is accompanied with the emergence of superconductivity in the collapsed phase. Recently, a very similar structural phase transition was discovered in LaFe$_2$As$_2$, but in contrast to CaFe$_2$As$_2$, superconductivity was only observed in the uncollapsed phase, not the collapsed phase. Previous studies have attributed this puzzling observation to the differences in the two materials' band coherence, orbital occupation, and Fermi surface topology. Here, we present a comparative study of LaFe$_2$As$_2$ and CaFe$_2$As$_2$ using the DFT+DMFT method. Surprisingly, we find that although La appears to have a valence higher than Ca, the doped one electron actually primarily resides on the La site. This leads to almost the same total Fe-$3d$ occupancy and electronic correlation strength as well as similar Lifshiftz transition in the Fermi surface topology for the two materials. In addition, we show that the two materials in both structural phases belong to the category of Hund's metals. Our results indicate that the electronic structures of LaFe$_2$As$_2$ and CaFe$_2$As$_2$ are not too different, which further suggest that superconductivity might also be induced in the collapsed phase of LaFe$_2$As$_2$ under similar non-hydrostatic conditions as for CaFe$_2$As$_2$.
\end{abstract}
%============== End abstract ==================%

%\keywords{supeconductivity, DFT+DMFT, correlation}

\maketitle

%
%============== Introduction ==================%
{\color{blue}\textit{Introduction.}} Electronic correlations, Fermi-surface topology and magnetism play important roles in unconventional iron-based superconductor materials~\cite{Kamihara.2008,Paglione.2010,Johnston.2010,Stewart.2011,Si.2016}. It is extremely intriguing that all three aspects can be dramatically changed during certain structural phase transitions. A prominent example is CaFe$_2$As$_2$ (Ca122), which exhibits a pressure-induced ``uncollapsed'' tetragonal (UT) to ``collapsed'' tetragonal (CT) structural phase transition~\cite{Ni:2008,Kreyssig:2008,Saha:2012,Bud'ko.2016,Yildirim.2008,Colonna.2011}.
At ambient conditions, Ca122 is in the UT phase with the ThCr$_2$Si$_2$-type lattice structure, and directly lowering the temperature below 170 K drives it into an orthorhombic phase, with a concomitant transition from paramagnetic to antiferromagnetic phase~\cite{Goldman.2008,Jeffries.2012}. Interestingly, upon applying a pressure, the magnetism can be suppressed, and the UT phase transits into the iso-structural CT phase which maintains the same crystal symmetry $I4/mmm$ but has a strong reduction of the $c$-axis by $\sim 8$\%~\cite{Goldman.2009,Pratt.2009,Kreyssig:2008}. More importantly, superconductivity was observed in the CT phase with $T_c$ of 12 K at 0.3 GPa~\cite{Torikachvili:2008,Jeffries.2012,Zhao.2015,Chen:2016}.

Recently, a similar UT to CT phase transition was reported in LaFe$_2$As$_2$ (La122)~\cite{Iyo:2019}. However, unlike Ca122, superconductivity (with $T_c\sim 12$ K) was observed in the UT phase, but not in the CT phase. The puzzle of the absence of superconductivity in CT-La122 has attracted significant attention~\cite{Mazin:2019,Usui:2019}. Formally, La$^{3+}$ has a higher valence than Ca$^{2+}$, hence one may tend to attribute the difference to the different doping levels of Fe $d$-orbitals, namely, La122 may be viewed as 0.5 electron doping from Ca122 (Fe$^{2+}$), resulting a nominal oxidation state with Fe$^{1.5+}$~\cite{Iyo:2019}. Particularly, in Ref.~\cite{Acharya:2020}, via a comparative study of La122 and Ca122 using \emph{ab initio} {QSGW}+DMFT techniques, it was found that CT-La122 exhibits much enhanced coherence of its electronic structure than CT-Ca122, and this distinction was argued to underly the absence/presence of superconductivity in CT-La122/Ca122. However, it is noted that the $c$ lattice parameter used for CT-Ca122 in Ref.~\cite{Acharya:2020} is only 1\% smaller than UT-Ca122, larger than the experimental value by $7\sim 8\%$~\cite{Saha:2012,Chen:2016}. This is a sizable difference. Hence, when taking experimental lattice parameters,
whether the comparison result still holds remains an open question.

In this work, we address this question by a comparative study of the electronic structures in both UT and CT phases of La122 and Ca122, using the DFT+DMFT method with an ``exact'' double-counting scheme. The key finding is that despite the different nominal valences of La and Ca,  La122 and Ca122 are \emph{not} too different in terms of their normal-state electronic structures. Particularly, the distinction between CT-La122 and CT-Ca122 regarding the coherence is not observed here. We show that both material systems have similar electronic correlation strengths and similar Lifshitz transitions of Fermi surface topology. These observations can be understood by noting that the extra electron doped by La in fact is largely retained at the La site rather than moving to the Fe-As layers that dominate the low-energy physics and (possible) superconductivity. In addition, we show that both materials belong to the category of Hund's metals where the Hund's coupling dominates the correlation effects~\cite{Werner.2008,Medici.2011,Medici.2011gs,Yin.2011oi,Georges.2013,Deng:2016,Haule:2009}. Since it was reported that superconductivity in CT-Ca122 is very sensitive to pressure and occurs only under non-hydrostatic conditions~\cite{Yu:2009}, our result implies that superconductivity may also be induced in CT-La122 under similar conditions.
%============== End introduction ==================%

\begin{table}
	\caption{\label{tab:struct} Lattice parameters $a$, $c$ and the As-Fe-As angles $\alpha_{\text{As-Fe-As}}$ used in this work~\cite{Iyo:2019,Saha:2012}. }
	\begin{ruledtabular}
	\begin{tabular}{cccc}
		\multirow{2}{*}{\ } &
		\multicolumn{3}{c}{Lattice Parameters} \\
		& $a$ (\AA) & $c$ (\AA) & $\alpha_{\text{As-Fe-As}}$ \\
		\hline
		UT-La122  & 3.938 & 11.732 & 110.8$\degree$  \\
		CT-La122  & 4.004 & 11.014 & 118.1$\degree$  \\
		\hline
		UT-Ca122  & 3.903 & 11.591 & 110.6$\degree$  \\
		CT-Ca122  & 3.982 & 10.684 & 115.9$\degree$
	\end{tabular}
	\end{ruledtabular}
\end{table}

%============== Method ==================%
{\color{blue}\textit{Methods.}} We perform fully charge self-consistent DFT+DMFT calculations using the {EDMFTF} software package~\cite{Haule:2010}, based on the full-potential linear augmented plane-wave method implemented in  the {WIEN2k} code~\cite{wien2k,Blaha.2020}.
A rotationally invariant form of local on-site Coulomb interaction Hamiltonian parameterized by Hubbard $U$ and Hund's coupling $J_H$ is applied on all five Fe-$3d$ orbitals. We choose $U=5.0$ eV and $J_H=0.8$ eV in this work, which are typical values for iron based superconductors~\cite{Yin.2011oi,ZPYin:2012,Mandal:2014}. The impurity problem is solved by the hybridization expansion version of the continuous-time quantum Monte Carlo solver~\cite{Gull.2011}.
We choose an ``exact'' double counting scheme developed by Haule~\cite{Haule:2015}, which eliminates the double counting issue in correlated materials. The self-energy on real frequency is obtained by the analytical continuation method of maximum entropy~\cite{Haule:2010}.
We use the experimental lattice parameters for both of the UT and CT phases of La122~\cite{Iyo:2019} and Ca122~\cite{Saha:2012}, which are listed in Table~\ref{tab:struct} (also see Supplemental Material (SM)~\cite{suppl}). The effective mass enhancement by correlations is defined by $m^*/m_{\text{DFT}} = 1/\mathcal{Z}$, where $\mathcal{Z}$ is the quasi-particle weight. We directly obtain $\mathcal{Z}=1-\frac{\partial\text{Im}\Sigma(\mathrm{i}\omega_n)}{\partial\omega_n}\big|_{\omega_n\rightarrow 0^+}$  from the self-energy on Matsubara frequencies, to avoid large error-bar in analytic continuations. More computation details are presented in SM~\cite{suppl}.
%============== End method ==================%

%==== figure =============================%
\begin{figure}
	\centering
	\includegraphics[width = 0.5\textwidth]{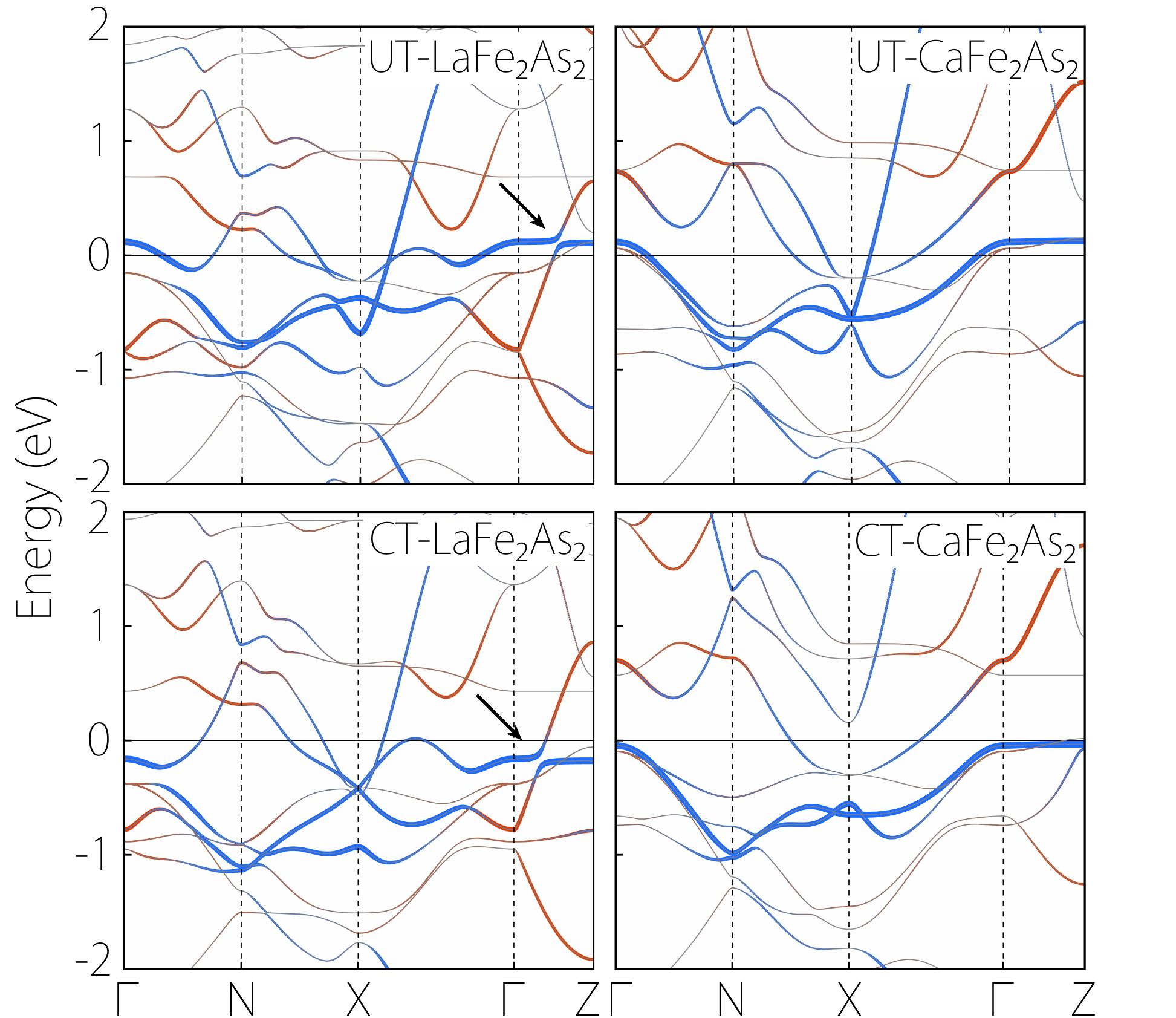}
	\caption{The DFT band structures for UT and CT phases of La122 and Ca122. Fat bands with Fe-$3d_{xy}$ and La-$5d$/Ca-$3d$ characters are shown in blue and red colors, respectively. The black arrows indicate the hybridization gaps between La-$5d$ and Fe-$3d$ bands.}
	\label{fig:ldaband}
\end{figure}
%=== end figure ==========================%

%==== Table =============================%
\begin{table*}
	\caption{\label{tab:correlation} (left) Total occupancy numbers of Fe-$3d$, La-$5d$/Ca-$3d$ and As-$4p$ orbitals, obtained from DFT and DFT+DMFT calculations at $T=300$ K. (middle) Orbital-resolved effective mass-enhancement $m^*/m_{\text{DFT}}=1/\mathcal{Z}$, and (right) the imaginary part of self-energy at zero frequency $-\text{Im}\Sigma(\mathrm{i}0^+)$ obtained from DFT+DMFT calculations at $T=300$ K.}
	\begin{ruledtabular}
	\begin{tabular}{cccc|ccc|cccc|cccc}
		\multirow{2}{*}{\ } &
		\multicolumn{3}{c|}{Occupancy by DFT} & \multicolumn{3}{c|}{Occupancy by DMFT} & \multicolumn{4}{c|}{$m^{*}/m_{\text{DFT}}=1/\mathcal{Z}$} & \multicolumn{4}{c}{$-\text{Im}\Sigma(\mathrm{i}0^+)$ (meV)} \\
		 & Fe-$3d$ & La/Ca-$d$ & As-$4p$ & Fe-$3d$ & La/Ca-$d$ & As-$4p$ & $d_{z^2}$ & $d_{x^2-y^2}$ & $d_{xz}/d_{yz}$ & $d_{xy}$ & $d_{z^2}$ & $d_{x^2-y^2}$ & $d_{xz}/d_{yz}$ & $d_{xy}$ \\
		\hline
		UT-La122  & 6.516 & 0.968 & 3.066 & 6.524 & 0.983 & 3.131 & 1.92 & 1.94 & 2.02 & 2.09 &  9.47 &  3.35 & 18.32 & 31.68  \\
		CT-La122  & 6.489 & 0.973 & 3.102 & 6.520 & 0.977 & 3.089 & 1.64 & 1.75 & 1.60 & 1.49 &  4.29 &  2.95 & 11.55 & 11.74  \\
		\hline
		UT-Ca122  & 6.507 & 0.383 & 3.056 & 6.546 & 0.385 & 3.078 & 1.83 & 1.80 & 2.00 & 1.94 &  4.00 &  2.93 & 15.33 & 17.18  \\
		CT-Ca122  & 6.515 & 0.394 & 2.993 & 6.559 & 0.396 & 3.016 & 1.65 & 1.75 & 1.79 & 1.60 &  0.86 &  0.18 &  5.24 &  4.47
	\end{tabular}
	\end{ruledtabular}
\end{table*}
%=== end table ==========================%

%============== Occupation ==================%
{\color{blue}\textit{Orbital-resolved occupancy numbers.}} Table~\ref{tab:correlation} shows the normalized total occupancy numbers of Fe-$3d$, La-$5d$/Ca-$3d$ and As-$4p$ orbitals, which include the states both inside atomic sphere and the interstitial region~\cite{Haule.2008}, obtained from our DFT and DFT+DMFT calculations. The most significant observation is that although La122 was previously considered as one electron doping from Ca122~\cite{Usui:2019}, with nominal oxidation state of Fe$^{1.5+}$, our calculations (both DFT and DFT+DMFT) show almost the same Fe-$3d$ occupancy $\sim6.5$ for both La122 and Ca122. (The extra 0.5 electrons on iron are due to strong Fe-$3d$ and As-$4p$ hybridizations.) Meanwhile, the La-$5d$ occupancy is about 1.0 in both UT and CT phases.  The As-$4p$ occupancies are also found to be very close in both materials ($\sim3.1$), which indicates their similar strengths of $3d$-$4p$ hybridizations. This is also reflected in the partial DOS presented in SM~\cite{suppl}.

These results clearly show that, from Ca122 to La122, the extra one electron brought in by La does not go to Fe-As layers (especially the Fe-$3d$ orbitals), instead it resides on the La site, which leads to an unusual oxidation state close to $+2$, not $+3$ for La.
This is related to the strong hybridizations between the very broad La-$5d$/$6s$ states and the Fe-$3d$ states, similar to the situation in the newly discovered infinite-layer nickelate superconductor, Nd$_{1-x}$Sr$_x$NiO$_2$, where the Nd-$5d$/$6s$ states also strongly hybridize with Ni-$3d$ states and contribute to the Fermi surface~\cite{Li.2019qa,Hepting.2020,Wang.20208l,Petocchi.2020}.
In Fig.~\ref{fig:ldaband}, we plot the DFT fat bands projected to the Fe-$3d_{xy}$ orbital (blue) and the La-$5d$/Ca-$3d$ orbitals (red). Indeed, we find substantial La-$5d$ states below the Fermi level for La122, whereas negligible Ca-$3d$ component is found in the similar range for Ca122.
Notably, such hybridization between Fe-$3d$ and La-$5d$/$6s$ states induces gaps in the $3d_{xy}$ band near the $\Gamma$ point in La122 (marked by the arrows in Fig.~\ref{fig:ldaband}), which is a major difference between band structures of La122 and Ca122.
The occupancy results for Ca122 obtained here agree well with previous calculations~\cite{Diehl.2014}.

Since the low-energy physics of these materials are mainly determined by the Fe-As layers, the similar total Fe-$3d$ occupancies should lead to their similar properties. Particularly, the two materials are expected to have similar electronic correlation strength and similar Lifshitz transition of Fermi surface across the UT-CT phase transition, as we demonstrate below.

%============== End occupation ==================%

% %==== figure =============================%
\begin{figure}
	\centering
	\includegraphics[width = 0.5\textwidth]{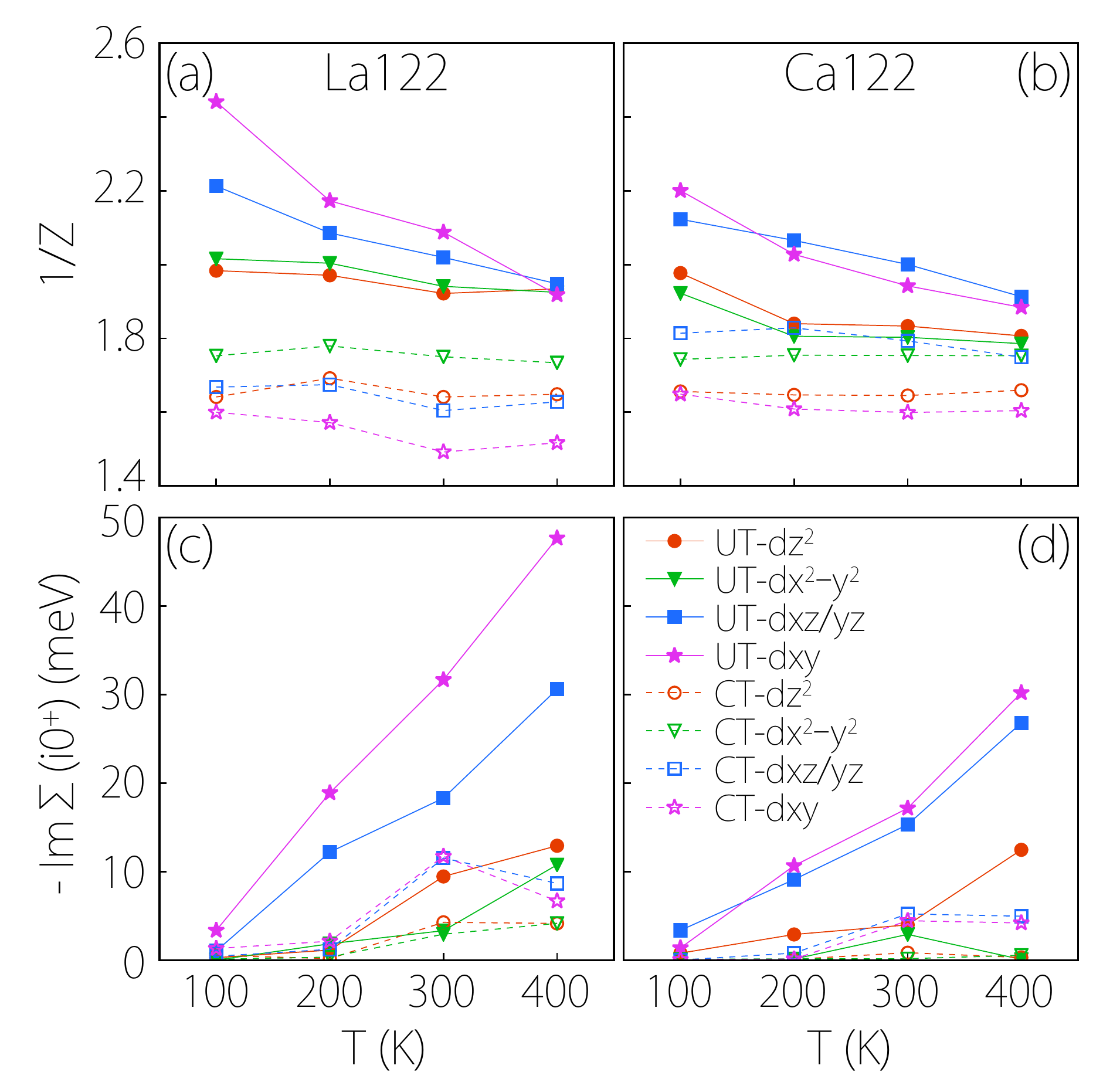}
	\caption{The variations of (a,b) orbital-resolved effective mass enhancement $m^*/m_{\text{DFT}} = 1/\mathcal{Z}$ and (c,d) the effective quasi-particle scattering rate $-\text{Im}\Sigma(\mathrm{i}0^+)$ for Fe-$3d$ orbitals as functions of temperature. The left (right) column is for La122 (Ca122). }
	\label{fig:hund}
\end{figure}
% %=== end figure ==========================%

%============== Correlations ==================%
{\color{blue}\textit{Electronic correlations.}} We then investigate the correlation effects in both UT and CT phases of La122, and make a comparison with Ca122.

The calculated orbital-resolved enhancement of effective mass $m^*/m_{\text{DFT}}$ and the quasi-particle scattering rate $-\text{Im} \Sigma(i0^{+})$ due to correlation effects at room temperature ($300$ K) are presented in Table~\ref{tab:correlation}(the imaginary part of Matsubara self-energy at room temperature are detailed in SM~\cite{suppl}). Let's first look at La122. In the UT phase, the mass enhancement ranges from 2.09 to 1.92, with $d_{xy}$ ($d_{z^2}$) as the most (least) correlated orbital. Especially, the $d_{xy}$ orbital shows much larger effective quasi-particle scattering rate than other orbitals. 
When transiting into the CT phase, the correlation effects are clearly weakened, which is reflected in the increased band width from UT to CT phase. 
In CT-La122, The mass enhancement ranges from 1.75 to 1.49, but now with the $d_{x^2-y^2}$ ($d_{xy}$) as the most (least) correlated orbital. The scattering rate decreases significantly across the phase transition. For the $d_{xy}$ orbital, the value decreases from 31.68 to 18.32. This is primarily due to that the $3d_{xy}$ bands are pushed below the Fermi level in the CT phase (see Fig.~\ref{fig:ldaband}), with substantial increase of the occupancy away from half filling.

In comparison, the results for Ca122 exhibit the same trend of variation in the UT-CT phase transition as expected from their similar total occupancy of Fe-$3d$ orbitals, which is also observed in the optical conductivity results~\cite{Xing.2016}. 
The mass enhancement of Ca122 in Table~\ref{tab:correlation} are consistent with experimental~\cite{Coldea.2009} and previous DFT+DMFT results~\cite{Diehl.2014,Mandal:2014}.
Particularly, the bands of CT-La122 (UT-La122) show similar coherence as CT-Ca122 (UT-Ca122). This differs from the result in Ref.~\cite{Acharya:2020}, in which a much less coherence was observed for CT-Ca122 (compared to CT-La122) as a larger $c$ lattice parameter was used.
%The correlation strengths of Ca122 found here are also consistent with previous DFT+DMFT works~\cite{Mandal:2014,Diehl.2014}.

We further study the temperature dependence of the mass enhancement and the quasi-particle scattering rate. The results are shown in Fig.~\ref{fig:hund}. One observes that the mass enhancement decreases and the scattering rate increases significantly with increasing temperature, for both UT and CT phases of the two materials. Especially, very large scattering rates appear at high temperatures. These are typical Hund's metal behaviors~\cite{Yin.2011oi,Georges.2013,Deng:2016,Haule:2009}.
We have also verified that these values would be strongly suppressed in the absence of Hund's coupling~\cite{suppl}.

%==== figure =============================%
\begin{figure}
	\centering
	\includegraphics[width = 0.5\textwidth]{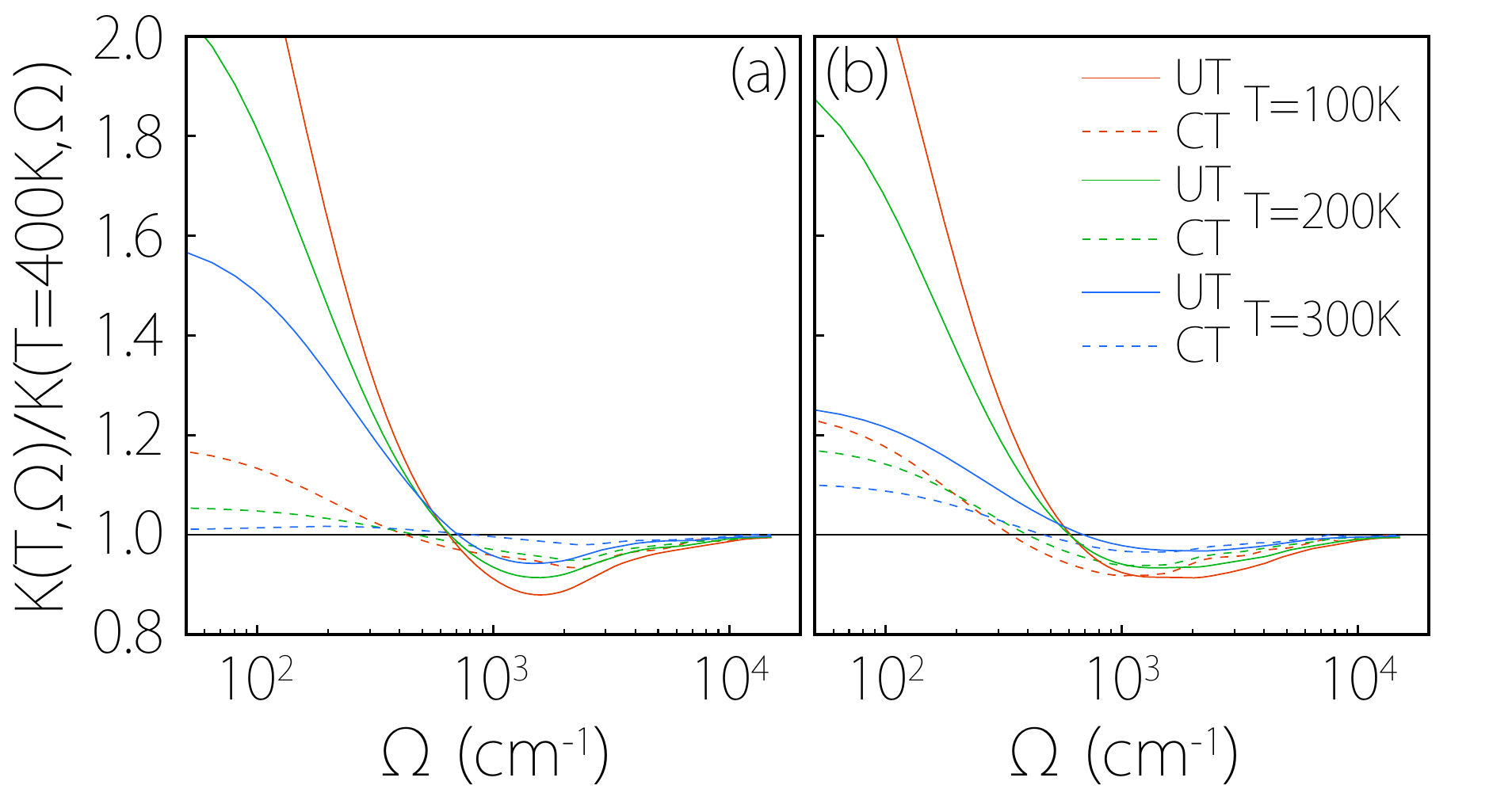}
	\caption{The ratio of SW, $K(T,\Omega)/K(400 K,\Omega)$, as a function of the cutoff frequency $\Omega$ for (a) La122 and (b) Ca122.}
	\label{fig:optic}
\end{figure}
%=== end figure ==========================%

The spectral weight (SW) transfer in the optical conductivity $\sigma_1(T, \omega)$ is considered as a hallmark of correlated multi-orbital physics~\cite{Qazilbash.2008,Wang.2012,Schafgans:2012,Deng:2016,Corasaniti.2020}.
Here, we calculate the optical conductivity (see SM~\cite{suppl}) and investigate the integrated SW:
\begin{equation}
	K(T, \Omega) = \frac{Z_0}{\pi^2} \int^\Omega_0{\sigma_1(T, \omega)\mathrm{d}\omega},
\end{equation}
where $Z_0=376.73\ \Omega$ is the impedance of free space. In Fig.~\ref{fig:optic}, we plot the ratio $K(T,\Omega)/K(400\ \text{K}, \Omega)$ as a function of the cutoff frequency $\Omega$. One observes that in each phase, the results for the two materials are similar and exhibit the same trend of SW transfer: With decreasing temperature, the SW transfers from $\sim$1000 cm$^{-1}$ of the mid infrared (MIR) range to both far infrared (FIR) and near infrared (NIR) ranges. The transfer to FIR represents the Drude component narrowing when temperature decreases, which is a common phenomenon for metals.
On the other hand, the transfer from MIR to NIR could be explained by the correlation effects due to Hund's coupling~\cite{Schafgans:2012}, which suppresses the spin fluctuation to form high spin state on the Fe ion (the atomic configuration is detailed in SM~\cite{suppl}). In addition, the SW ratio recovers to $\approx 1$ at about 12000 cm$^{-1}$, at which the energy scale of spectral-weight transfer is defined. This scale is about the same size of Hund's coupling $J$ in our calculation. As a comparison, we note that
in the metallic phase of Mott-Hubbard systems, the SW transfer direction is from high to low energies into the Drude component with decreasing temperature~\cite{Qazilbash.2008}. And in that case, the energy scale is associated with the Hubbard $U$, which is much larger than the Hund's coupling.
%============== End Correlations ==================%

%============== FS ==================%

%==== figure =============================%
\begin{figure*}[!ht]
	\centering
	\includegraphics[width = 0.98\textwidth]{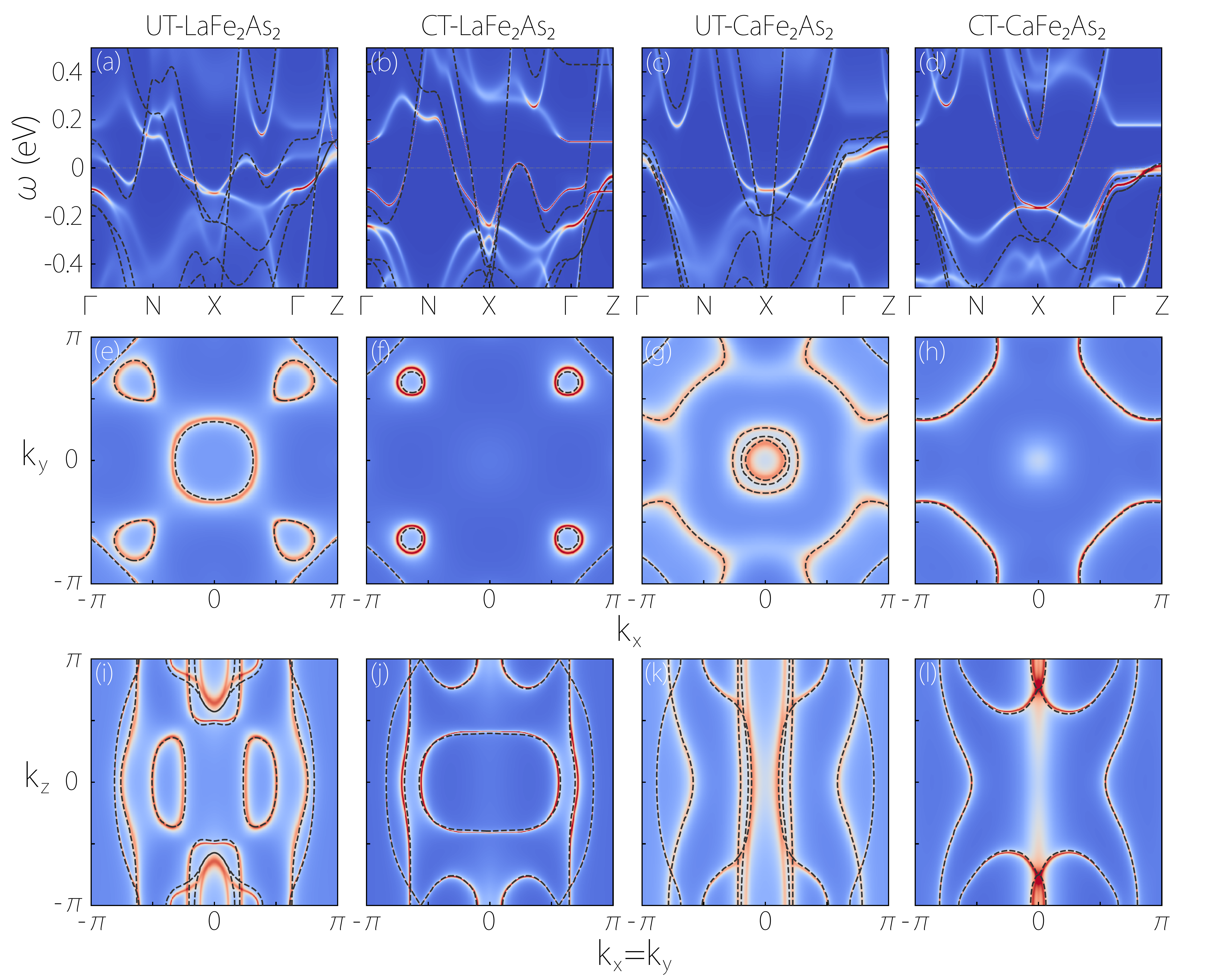}
	\caption{The correlated electronic structures obtained from the DFT+DMFT calculations at $T=300$ K. The columns from left to right are for UT-La122, CT-La122, UT-Ca122 and CT-Ca122, respectively. (a)-(d) Momentum resolved spectral function $\mathcal{A}(k,\omega)$. (e)-(h) Fermi surface in the (001) plane. (i)-(l) Fermi surface in the (110) plane. The corresponding DFT results are shown in dashed lines for reference.}
	\label{fig:akw}
\end{figure*}
%=== end figure ==========================%
{\color{blue}\textit{Fermi surfaces.}} The momentum-resolved spectral functions $\mathcal{A}(k,\omega)$ from our DFT+DMFT calculations at $T=300$ K are plotted in Fig.~\ref{fig:akw}(a-d).
The quasi-particle bands near the Fermi level are obviously renormalized compared with the LDA bands (black dashed lines). The quasi-particle bandwidth is narrowed to about half of the LDA bands, which is consistent with renormalization factor $\mathcal{Z}$ discussed above. The correlated electronic structure undergoes obvious incoherence to coherence crossover during the UT-to-CT phase transition in both La122 and Ca122. Again, one can observe that CT-La122 and CT-Ca122 have similar coherence. In addition, we find that the all spectra become much more coherent when ignoring the Hund's coupling~\cite{suppl}, manifesting features of Hund's metals~\cite{Georges.2013}.

In Fig.~\ref{fig:akw}(e-l), we plot the corresponding Fermi surfaces in the $k_z=0$ plane and in the $k_x=k_y$ plane.
Overall, compared to the DFT results, the Fermi surface topology is not significantly modified by the correlation effects.
For UT-La122, there is one hole Fermi surface with $d_{xy}$ character in the $k_z=0$ plane around the $\Gamma$ point (Fig.~\ref{fig:akw}(e)), while for UT-Ca122, there are three hole Fermi surfaces (Fig.\ref{fig:akw}(g), the outer one is of $d_{xy}$ character and the inner two are of $d_{xz}$/$d_{yz}$ character~\cite{Diehl.2014}).
In the $k_x=k_y$ plane, UT-La122 has another Fermi surface with La-$5d$ character around $\Gamma$ (Fig.~\ref{fig:akw}(i)), while it disappears in UT-Ca122 because of the negligible Ca-$3d$ occupancy (Fig.~\ref{fig:akw}(k)).
Across the UT-CT phase transition, the most obvious change is that the conduction band at $\Gamma$ point in UT phase is pushed down below the Fermi level in the CT phase, which occurs for both materials (see also Fig.~\ref{fig:ldaband}). As a result, the hole Fermi surfaces around $\Gamma$ in UT phase of both materials disappear in the CT phase, corresponding to a Lifshitz transition of Fermi surface topology.
%============== End FS ==================%

%============== Conclusion ==================%
{\color{blue}\textit{Discussion.}} Our result shows that in either UT or CT phase, La122 and Ca122 are similar in terms of qualitative band features, correlation strength, band coherence, SW transfer, and Fermi surface topology. All these are connected with the unusual oxidation state of La which results in their similar Fe (and As) orbital occupation numbers. Particularly, we do not observe a marked difference in band coherence and Fermi surface topology between CT-La122 and CT-Ca122, indicating that these should not be primary factors for their possibly different superconducting behaviors. In fact, it was noted that superconductivity in CT-Ca122 is very sensitive to pressure and only occurs under non-hydrostatic conditions~\cite{Torikachvili:2008,Yu:2009}. Based on their similarity, the same may also be true for CT-La122. We suggest further experiments to test out possible superconductivity in CT-La122 with different pressure conditions.

Our result also indicates that both phases of La122 (and Ca122) are Hund's metals in the normal state. Recently, the discovery of superconductivity in Nd$_{1-x}$Sr$_x$NiO$_2$ attracted great interest~\cite{Li.2019qa}. There, it was shown that doped NdNiO$_2$ is a multi-orbital Hund's metal, and Ni-$3d$ occupancies are very similar in both LaNiO$_2$ and SrNiO$_2$ compounds~\cite{Wang.20208l,Petocchi.2020}. Those results are very similar to our findings here. Therefore, this study may also help to shed light on the common features of Hund's superconductors.
%============== End conclusion ==================%

%============== Acknowledgment ==================%
%\textit{Acknowledgment.}---We thank Hu Miao for very helpful discussions. This work was supported by the National Natural Science Foundation of China (No. 11604273), and the Singapore Ministry of Education AcRF Tier 2 (MOE2019-T2-1-001). Y. W. was supported by the U.S. Department of Energy, Office of Science, Basic Energy Sciences as a part of the Computational Materials Science Program through the Center for Computational Design of Functional Strongly Correlated Materials and Theoretical Spectroscopy.
%============== End acknowledgment ==================%

%\begin{footnotesize}
\bibliography{fe122_bib}
%\end{footnotesize}

\end{document}